\documentclass{elsart5p}

\usepackage{natbib}

\usepackage{amsmath}   
\usepackage{graphicx}  

\renewcommand{\vec}[1]{\mathsf{#1}}
\newcommand{\matr}[1]{\mathsf{#1}}
\newcommand{\lev}[2]{\mathsf{#1}_{#2}}
\newcommand{\rev}[2]{\mathsf{#1}_{#2}}
\newcommand{\scp}[2]{#1 \cdot #2}

\newcommand{\rmd}{\mathrm{d}}
\newcommand{\rme}{\mathrm{e}}

\begin{document}

\begin{frontmatter}

\title{The propagation of a cultural or biological trait by neutral
genetic drift in a subdivided population}

\author{R.\ A.\ Blythe}

\address{School of Physics, University of Edinburgh,
Mayfield Road, Edinburgh EH9 3JZ, UK}

\begin{abstract}
We study fixation probabilities and times as a consequence of neutral
genetic drift in subdivided populations, motivated by a model of the
cultural evolutionary process of language change that is described by
the same mathematics as the biological process.  We focus on the
growth of fixation times with the number of subpopulations, and
variation of fixation probabilities and times with initial
distributions of mutants.  A general formula for the fixation
probability for arbitrary initial condition is derived by extending a
duality relation between forwards- and backwards-time properties of
the model from a panmictic to a subdivided population.  From this we
obtain new formul\ae\, formally exact in the limit of extremely weak
migration, for the mean fixation time from an arbitrary initial
condition for Wright's island model, presenting two cases as examples.
For more general models of population subdivision, formul\ae\ are
introduced for an arbitrary number of mutants that are randomly
located, and a single mutant whose position is known.  These
formul\ae\ contain parameters that typically have to be obtained
numerically, a procedure we follow for two contrasting clustered
models.  These data suggest that variation of fixation time with the
initial condition is slight, but depends strongly on the nature of
subdivision.  In particular, we demonstrate conditions under which the
fixation time remains finite even in the limit of an infinite number
of demes.  In many cases---except this last where fixation in a finite
time is seen---the time to fixation is shown to be in precise
agreement with predictions from formul\ae\ for the asymptotic
effective population size.
\end{abstract}

\begin{keyword}
Random genetic drift \sep Population subdivision \sep Migration \sep
Effective population size \sep Coalescent \sep Cultural evolution
\end{keyword}

\end{frontmatter}

\section{Introduction}

Genetic drift is a generic term for fluctuations in allele frequencies
that arise by sampling a finite population to produce offspring in the
next generation.  In the absence of mutation, these fluctuations can
lead to extinction of some alleles, ultimately causing one allele to
fix.  One can therefore ask whether it is feasible for some trait to
propagate across an entire population by neutral genetic drift alone.
In the simplest mathematical models, such as those due to
\citet{Fisher30}, \citet{Wright31} and \citet{Moran58}, individuals
from the entire population mate randomly and it is known that fixation
time (measured in units of the expected lifetime of one individual in
the population) increases linearly with the population size
\citep{KO69,CK70}.  This growth law calls into question the viability
of neutral genetic drift as a mechanism for population-level change,
on the grounds that changes in large populations are simply far too
slow.  An important question then arising is whether \emph{non-random}
mating---for example, that seen in subdivided populations---can reduce
fixation times in large populations.

This very same question has recently arisen in the context of the
cultural evolutionary phenomenon of language change, in which the unit
of variation is some aspect of spoken language, such as a vowel sound,
through which one can distinguish different dialects of the same
language, but which are otherwise functionally equivalent.  A simple,
agent-based model of language reception and reproduction
\citep{BBCM06}, has a mathematical description that coincides with
that of neutral genetic drift in a subdivided population, although a
number of details of the underlying evolutionary processes are rather
different---in particular, the language does not evolve due to genetic
changes in the speakers, but rather the frequencies of linguistic
variants in the population of utterances change over time in a manner
akin to genetic drift.  The key points here, however, are that
fixation of an allele corresponds to a linguistic innovation being
adopted as a community's convention, and that interactions between
speakers map to migrations between large subpopulations.  Thus here
even a linear growth in fixation time with the number of speakers
(each one corresponding to a single subpopulation) is untenable: a
change that becomes established in a small clique in a few days would
then require several hundred years to propagate across a
modestly-sized society.  To see why this is a problem, one should note
that this long timescale roughly corresponds with the age of language
itself \citep{CK03}, whilst society-wide language change has been seen
to occur within one or two human generations, for example in the case
of new dialect formation \citep{Gordon04,Trudgill04}.

It is therefore tempting to suggest that some form of selection, i.e.,
intrinsic fitness of some linguistic variants over others, is required
for an innovation to spread rapidly across an entire society.  To be
able to state this categorically, however, we must understand
quantitatively how fixation times and probabilities depend on the
network of interactions between speakers in the society (which, in the
biological interpretation of our model, corresponds to a set of
migration rates between geographically separated subpopulations) and
the overall size of that network (the number of subpopulations).  We
also need to explore how these fixation statistics vary with changes
in the initial condition (distribution of mutants across the total
population) since this is the information provided in the historical
data that we hope to use to infer the propagation mechanisms of
linguistic innovations.  In this work, we aim to develop this required
understanding.

A recurring theme in the considerable literature on genetic drift in
subdivided populations---this beginning with \citet{Wright31} and
continuing through to the present day \citep[see e.g][for recent
reviews]{CCB03,Wakeley05}---is that properties of fixation are not
strongly affected by such subdivision.  For example, when migration is
conservative, the probability a mutant allele fixes, whether
selectively advantageous or not, is the same as that as in an ideal
panmictic population with the same overall size
\citep{Maruyama70a,Slatkin81}.  In particular this means that no
matter where mutants are initially positioned, they fix with equal
probability.  For this state of affairs to change, one either has
either has to introduce additional processes---such as extinction and
recolonisation \citep{Barton93}---or relax the assumption of
conservative migration \citep{LHN05} in which case certain
geographical structures can result in an allele with even a small
selective advantage fixing with near certainty.  Whilst the dependence
of fixation probability on the initial mutant distribution can be
established rather easily (see below), the corresponding variation of
fixation time seems harder to obtain and progress in this direction
forms the bulk of the present work.

A continued study of the literature reveals that the concept of
effective population size---which again goes back to
\citet{Wright31}---has proved useful in characterising the overall
fixation timescale.  Essentially, the idea is that many properties of
neutral genetic drift in a subdivided population comprising in total
$N$ instances of a gene\footnote{It is traditional in the population genetics
literature to talk in terms of a number of diploid organisms, in which
case the number of genes $N$ is twice the number of organisms.
Since ploidy is irrelevant in the cultural evolutionary application,
we shall suppress these factors of two.} are the same as those for an
ideal panmictic population with an effective number of gene instances $N_e$.

One prominent definition of effective population size relates to the
change per generation in the variance of a mutant allele frequency
\citep{CK70}.  Let $x$ be the frequency of mutants present in a
population at some time, and $x'$ their frequency after a single
generation.  Then, the variance effective size can be defined as
$N_e=x(1-x)/{\rm Var}(x')$.  If this effective population size is
large, one can then use a diffusion approximation to model the
evolution of the distribution of mutant frequencies.  One can then
calculate the mean time to fixation to be
\begin{equation}
\label{tideal}
\tau \approx - 2N_e \frac{(1-x)\ln(1-x)}{x}
\end{equation}
as was shown in the classic work by \citet{KO69}.  Thus if one has the
variance effective population size for the kind of large, subdivided
population that is of interest in the present work, Eq.~(\ref{tideal})
can be used to estimate the fixation time.  In principle such an
effective population size could be extracted from the diffusion
approximation that applies to genetic drift in a subdivided
population.  However, one typically finds that the effective size
depends on the mutant allele frequency $x$ \citep[see
e.g.][]{RR03,Rousset03} which makes the resulting expressions
difficult to handle, unless additional approximations are made such as
that used by \citet{CW03}.  Instead, we exploit here the fact that
equilibrium values of genetic variables, such as the frequency of
mutants, are all approached exponentially with a common time constant
asymptotically \citep{WB97,Rousset04}.  This time constant can then be
used to define an effective population size.

One way to do this is to consider the change per generation in the
probability that two individuals sampled from the population are
identical by descent, i.e., share a common ancestor
\citep{WB97,Rousset04}.  Let us recapitulate here the derivation of
the formula for asymptotic effective size given by \citet{Rousset04}
for the model of population subdivision that will be used throughout
this work.  This model comprises $L$ demes, each having a fixed
subpopulation size $N_i$.  After a generation of reproduction and
migration, the probability that an individual in deme $i$ is an
offspring of a parent in deme $j$ is $\mu_{ij}$, hence
$\sum_{j}\mu_{ij}=1$ for all $i$.  Given the set of probabilities
$F_{ij}$ that two individuals, one sampled from deme $i$, one from
deme $j$, share a common ancestor, one finds one generation later that
the corresponding set of probabilities are given by
\begin{equation}
\label{F'}
F_{ij}' = \sum_{k,\ell} \mu_{ik} \mu_{j\ell} \left( F_{k\ell} +
\frac{1-F_{kk}}{N_k} \delta_{k, \ell} \right) \;.
\end{equation}
This form arises because if the parents of the two sampled individuals
came from different demes $k$ and $\ell$, they are identical by
descent with probability $F_{k \ell}$, whereas if the parents are from
the same deme $k$ they have a probability $1/N_k$ of being the same
individual (and hence genetically identical in the absence of
mutation) whereas otherwise the probability they are identical is
$(1-1/N_k)F_{kk}$.

To simplify this expression, we introduce the important quantity
$Q_i^\ast$ which is defined as follows.  Consider some subset of the
present-day population, and trace its ancestry backwards in time until
a single common ancestor is found.  Now, if one keeps tracing
backwards in time, this ancestor will hop from deme to deme and
eventually the probability that it is to be found in deme $i$
approaches the stationary value $Q_i^\ast$.  Since a hop from deme $i$
to $j$ occurs with a probability $\mu_{ij}$ per generation, these
stationary probabilities satisfy the equation
\begin{equation}
\label{ssmu}
\sum_{i} Q^\ast_i \mu_{ij} = Q^\ast_j \;.
\end{equation}
Multiplying both sides of (\ref{F'}) by $Q^\ast_i Q^\ast_j$ and
summing over all $i$ and $j$, one then finds that
\begin{equation}
\Delta \tilde{F} \equiv \tilde{F}' - \tilde{F} = \sum_{i=1}^{L}
(Q^\ast_i)^2 \frac{1 - F_{ii}}{N_i} \;,
\end{equation}
where $\tilde{F} = \sum_{ij} Q^\ast_i Q^\ast_j F_{ij}$.  As
$t\to\infty$, the probability that any pair of individuals share a
common ancestor approaches unity under neutral evolution.  As this
limit of infinite time is approached, one can unambiguously define an
asymptotic effective size of the subdivided population through
\begin{equation}
\label{Ne}
\frac{1}{N_e} = \lim_{t \to \infty} \frac{\Delta
\tilde{F}}{1-\tilde{F}} = \lim_{t\to\infty} \frac{\sum_{i=1}^{L}
\frac{1}{N_i} (Q_i^\ast)^2 (1-F_{ii})}{\sum_{i=1}^{L} \sum_{j=1}^{L}
Q_i^\ast Q_j^\ast (1-F_{ij})}
\end{equation}
where $t$ is the number of generations that have elapsed since
imposing some initial condition.  One way to interpret this equation
is as an average of coalescence rates $1/N_i$ of two lineages present
in the same deme $i$ in the limit $t\to\infty$ \emph{given} that they
have not previously coalesced \citep{Rousset04}.

If migration is a fast process, one can envisage that the identity
probability $F_{ij}$ is roughly the same for all pairs of demes $i$
and $j$.  Under such conditions, the formula simplifies to
\begin{equation}
\label{NeNag}
\frac{1}{N_e} = \sum_{i=1}^{L} \frac{(Q_i^\ast)^2}{N_i} \;.
\end{equation}
This formula was obtained by an independent means by
\citet{Nagylaki80}, and was shown to be valid in the limit of infinite
deme sizes when migration rates per generation satisfy $\lim_{N_i \to
\infty} \mu_{ij} N_i = \infty$, an expression that defines a fast
migration limit \citep[see][for a fuller discussion of the various
limits that are of interest]{Wakeley05}.  For intermediate migration
rates, \citet{WB97} presented a formula in which the denominator of
(\ref{Ne}) is replaced by $1-\bar{F}$ where $\bar{F}$ is the identity
probability averaged uniformly over all pairs of individuals in the
population.

In this work we are almost exclusively interested in the slow
migration limit, where deme sizes again again approach infinity, but
one has $\lim_{N_i \to \infty} \mu_{ij} N_i < \infty$.  There are
several reasons for this.  Firstly, this is the limit that naturally
arose in the model of language change previously described
\citep{BBCM06}.  In a population genetics context this limit is also
relevant when the subpopulation sizes are large and one is interested
in the evolution over a number of generations that is of the order of
these subpopulation sizes.  Finally, it is a limit in which any
deviation from panmictic behaviour---such as sensitivity to an initial
distribution of mutants---would be most likely to arise, and thus
worth exploring from a more theoretical point of view \citep[see,
e.g.,][for a discussion of the different phenomenology in the fast and
slow migration limits]{Slatkin81}.  We remark that this is not the
only slow migration limit that can be taken: one can also take
$\mu_{ij}\sim \mu \to0$ at fixed population sizes and measure time in
units of $1/\mu$ \citep[see, e.g.,][]{Wakeley04}.  For clarity, we
reiterate that throughout most of this work, we deal with a slow
migration limit in which deme population sizes $N_i$ are infinite, and
migration rates are inversely proportional to this size.  Furthermore,
for reasons to be discussed in the next section, we will most often be
dealing with the extreme case where the coefficients are vanishingly
small.

In this limit, it is unclear whether one can simply use the expression
(\ref{Ne}) for the effective size that appears in Eq.~(\ref{tideal})
for the fixation time.  Tracing the ancestry of a state of fixation
backwards in time, the extreme slow migration regime could lead to a
situation in which the majority of the relevant coalesence events
occur long before the time at which the mean coalescence rate given by
the right-hand size of (\ref{Ne}) is reached.  We also remark that it
has also been argued that only when lineages are able to equilibrate
through a rapid migration process between coalescent events, is it
appropriate to characterise the coalescence rates by a single
effective population size \citep{NK02,SKKLN05}.  This fact is
particularly relevant when considering a final state of fixation,
since one must trace the ancestry of the entire population.  Finally,
whether or not the effective population size turns out to provide good
estimates of fixation times in the slow migration limit, it does not
provide a framework for understanding how fixation times vary with the
initial distribution of mutants.

To assess the utility of the asymptotic effective population size in
characterising fixation timescales, and to investigate the effect of
varying the initial condition, we develop a formalism within which the
spatial distribution of mutants as a function of time can (in
principle, at least) be calculated exactly given any initial
distribution and model of a subdivided population undergoing neutral
genetic drift.  Specialising to the case of a final state of fixation,
we can extract via Eq.~(\ref{tideal}) a measure of effective
population size and compare with its asymptotic counterpart using
Eq.~(\ref{Ne}).  As has been hinted above, the approach we use is
based on the backward time coalescent process \citep{Kingman82} that
has found many applications in the context of drift in subdivided
populations \citep[see,
e.g.,][]{Notohara90,Takahata91,NT93,DT95,Wakeley98,WilkinsonHerbots98,Notohara01,Wakeley01,CCB03,WL06}.
The main benefit of this approach is that lineages that do not
contribute to the final state of fixation are explicitly excluded from
the mathematical description.  However, a fixed initial condition does
not enter naturally in this description, and so some work is needed to
match this up with the distribution of ancestors of the present day
state of fixation.  As we have already remarked, population
subdivision is harder to handle in diffusion-equation-based
approaches, although the initial condition is more easily enforced and
extensions to alleles with a selective advantage can be treated more
straightforwardly \citep[see,
e.g.,][]{Maruyama70a,Slatkin81,Barton93,Whitlock02,CW03,RR03,WT04}.

In the next section we will present the details of the derivation of
the exact formula relating forward- and backward-time properties to
one another, and demonstrate the simplifications that occur in the
limit of extemely slow migration.  The rest of this work is then
devoted to consequences of this formula under different models of
subdivision.  It is worth at this early stage to fix the basic ideas
underlying the approach by considering the simple case of fixation
probability as a function of initial condition.  Consider therefore
the probability that a mutant allele fixes when initially a fraction
$\chi_i$ of the individuals in deme $i$ are mutants.  We denote this
probability $P^\ast(A)$, where at this stage $A$ is a shorthand for
the initial condition (we will define it more formally below).  If we
look infinitely far forward in time from this initial state, fixation
of either the mutant or the wild type is guaranteed to have occurred.
Then rewinding infinitely far from the final state back to the initial
state, we find a single ancestor in deme $i$ with probability
$Q_i^\ast$ as previously discussed.  Since the initial assignment of
mutants to individuals in the population is completely independent of
the ensuing population dynamics, this ancestor is a mutant with
probability $\chi_i$. Hence, the fixation probability is simply
\begin{equation}
\label{Past}
P^\ast(A) = \sum_{i=1}^{L} \chi_i Q_i^\ast \;.
\end{equation}
Thus to find the fixation probability from any initial condition, it
is enough to know the distribution $Q_i^\ast$ which is the stationary
distribution of the Markovian migration process running backwards in
time.  Since finding this stationary distribution is a fairly standard
problem in the theory of stochastic processes, we shall not consider
it in further detail here, other than to remark that through variation
in the migration rates, it is possible to construct models in which
$Q_i^\ast$ for some deme $i$ is close to one.  If selectively neutral
mutants initially completely occupy that deme they are then almost
guaranteed to spread to the whole population.  This is similar in
spirit to the phenomenon discussed by \citet{LHN05}, in which it was
seen that the spatial structure of the population could give rise to a
mutant fixing with near certainty.  The distinction is that in that
work, the mutant was considered to have a selective advantage and thus
that the initial location played only a minor role in the probability
of fixation.  Under neutral genetic drift, the situation is reversed:
the same mutation can fix or go extinct with near certainty depending
on where it occurs.  This will be demonstrated explicitly in
Section~\ref{singleseed} below.

Before this, in Section~\ref{ranic} we derive from the exact
formul\ae\ presented in Section~\ref{prelim} an expression that holds
for arbitrary population subdivision and gives the mean fixation time
from an initial condition in which mutants are randomly distributed
with an overall frequency $x$.  This would be the appropriate initial
condition to use, for example, when modelling a historical situation
for which initial mutant frequencies, but not their location, are
known.  We learn that the resulting expression involves the mean times
to particular coalescence events from the final state of fixation,
which need then to be obtained either analytically or by Monte Carlo
sampling.  One case in which the former is possible is Wright's island
model \citep{Wright31,Maruyama70b,Latter73}; furthermore, the formula
can be extended to an arbitrary initial condition.  Therefore, in
Section~\ref{wright} we are able to perform a number of exact
calculations for this very well-studied model, apparently for the
first time.  For example, we can consider in addition to the initial
condition that has mutants initially randomly scattered across all
islands with some overall mutant frequency $x$ the contrasting case
where the same number of mutants are all initially confined to the
smallest possible number of islands.  In genetics terms, these two
different initial conditions correspond to the two extreme values of
the inbreeding coefficient $F_{\rm ST}=0$ and $1$ respectively.  For
the island model, we show that the difference in fixation time for
these two initial conditions is short on the timescale of fixation
which grows linearly with the number of demes.  We further provide
evidence that fixation from any non-random initial condition is slower
than from a random initial condition.  For more general models, exact
calculations are much more difficult.  Nevertheless, we introduce a
formula for fixation time from a single mutation with known location
(a case of practical interest) to the statistics of the most recent
common ancestor of the whole population. This will be explained in
Section~\ref{singleseed}.  By combining analytical results with
numerical data from simulations we explore fixation times, and their
variation with initial condition, in two concrete and contrasting
models in which there demes fall into clusters within which migration
occurs at different rates than between demes from different clusters.
Our interest here is in how fixation times grow with increasing number
of demes, either by adding more demes to each cluster, or by adding
clusters of fixed size.  Whilst the former case has been considered in
a number of works \citep[such as][]{Wakeley01,WA01,WL06}, we obtain in
this work some evidence that in the latter case, the mean time to
fixation can remain \emph{finite} in the limit of an infinite number
of demes.  This effect---and the general growth law for the fixation
time in a subdivided population---is predicted by the effective size
formula (\ref{Ne}) in the slow migration limit.  In all cases other
than that in which fixation in a finite time is seen, the coefficient
of the growth law is also accurately predicted by (\ref{Ne}).  We
discuss possible origins for the discrepancy that remains, along with
implications of our findings for more plausible models of social
interaction and migration between biological populations in the
conclusion, Section~\ref{discon}.

\section{Fixation probability as a function of time}
\label{prelim}

We begin by deriving an expression for the distribution of mutants
that is valid for any model of population subdivision and arbitrary
initial condition.  We do however stipulate that under the population
dynamics that an individual's chances of leaving offspring in the next
generation are unaffected by whether it is a mutant or not (i.e.,
there is no selection), and that subpopulation sizes and migration
rates are constant in time.

As stated in the introduction, we consider a population that is
subdivided into $L$ demes with $N_i$ individuals in deme $i$.  The
initial condition, denoted $A$, is specified by the number of mutants
$a_i$ in each deme, $i=1,2,\ldots,L$.  Two distinct distributions are
central to this work.  The first is the probability $P(B|A;t)$ that
after $t$ generations of reproduction within and migration between
demes, all descendants of $A$ form the set $B$, i.e., a distribution
with precisely $b_i$ mutants in deme $i$.  The second is the
probability $Q(C|D;t)$ that all ancestors of a present-day sample $D$
formed $t$ generations previously the set $C$.  The quantities $d_i$
and $c_i$ are defined analogously to $a_i$ and $b_i$ for these sets.

These two distributions are connected by a duality relation that was
given by \citet{Mohle01} for the panmictic case, $L=1$.  The
generalisation to $L>1$ is found by recalling that---in the absence of
selection---the assignment of mutant alleles to individuals within the
population is a process independent of the population dynamics.  We
proceed by constructing equivalent expressions---one for $P$ and one
for $Q$---for the probability of the event that a set of individuals
$A$ contains all ancestors of some set of individuals $D$ present in
the population $t$ generations later.  To be clear, the set $A$ must
contain at least the ancestors of individuals in $D$, and may
additionally contain ancestors of individuals outside $D$, or
individuals that have no descendants at the later time; likewise,
descendants of $A$ may form a superset of the individuals in $D$.
Therefore, we obtain the desired probability by summing $P(B|A;t)$
over all sets of individuals $B$ that contain $D$; meanwhile
$Q(C|D;t)$ must be summed over all sets $C$ that are contained within
$A$.  In both cases, the sum must be weighted by the probability that
one randomly chosen set is contained within the other.  The expression
that results is
\begin{multline}
\label{dual}
\sum_{b_1=d_1}^{N_1} \sum_{b_2=d_2}^{N_2} \cdots \sum_{b_L=d_L}^{N_L}
  \prod_i \frac{{ b_i \choose d_i }}{{ N_i \choose d_i }} P(B|A; t)
  =\\ \sum_{c_1=0}^{a_1} \sum_{c_2=0}^{a_2} \cdots
  \sum_{c_L=0}^{a_L}\prod_i \frac{{ a_i \choose c_i }}{{ N_i \choose
  c_i }} Q(C|D; t) \;.
\end{multline}

In this work, we will find $Q(C|D;t)$, or quantities derived from it,
either analytically or numerically, and use this duality relation to
derive new formul\ae\ relating to fixation. This requires us to to
rearrange the implicit expression (\ref{dual}) so that the
forward-time probability $P$ (that relates to fixation statistics) is
given explicitly in terms of $Q$.  This is achieved by making use of
the identity
\begin{eqnarray}
\label{id}
\sum_{j} (-1)^{i+j} \frac{{ N \choose i}{N-i \choose j-i} {k \choose
j}}{{N \choose j}} &=& (-1)^i {k \choose i} \sum_{j=i}^{k} (-1)^j {k-i \choose j-i}\nonumber\\
 &=& \delta_{i,k} \;,
\end{eqnarray}
in which $\delta_{i,k}$ is the Kronecker delta symbol.  Here, the
first step is achieved by expanding the binomial coefficients in terms
of factorials, making some cancellations and recombining remaining
factorials as binomial coefficients.  Then, if $k<i$, the sum is zero
by definition; if $k>i$, the alternating binomial coefficients sum to
zero \citep[see e.g., formula (0.15.4) in][]{GR00}; if $k=i$ one can
easily verify that that the resulting expression equals unity, as
required.  Multiplying both sides of (\ref{dual}) by
\[
\prod_i \sum_{d_i} (-1)^{b'_i+d_i} {N_i \choose b'_i} {N - b'_i
  \choose d_i - b'_i}
\]
and summing over all $b_i$ and $d_i$ one finds
\begin{multline}
\label{PofQ}
P(B|A;t) = 
\sum_{c_1=0}^{a_1} \sum_{c_2=0}^{a_2} \cdots \sum_{c_L=0}^{a_L} 
\sum_{d_1=b_1}^{N_1} \sum_{d_2=b_2}^{N_2} \cdots \sum_{d_L=b_L}^{N_L} \\
\prod_i (-1)^{b_i+d_i} \frac{{N_i \choose b_i}
  {N_i - b_i \choose d_i - b_i} {a_i \choose c_i}}{ { N_i \choose c_i}
  } Q(C|D; t)
\end{multline}
after dropping the prime on the variables $b_i$ that remain. For a
final state of fixation, $b_i=N_i$ and this expression simplifies to
\begin{equation}
\label{PAt}
P(A; t) = \sum_{c_1=0}^{a_1} \sum_{c_2=0}^{a_2} \cdots \sum_{c_L=0}^{a_L}
\prod_i \frac{{ a_i \choose c_i }}{{ N_i \choose c_i }} Q(C; t) \;,
\end{equation}
in which we have introduced the notation $P(A;t)$ for the probability
of fixation from an initial distribution of mutants $A$, and $Q(C;t)$
for the distribution of ancestors of the entire population a time $t$
previously.  Although we are concerned only with fixation in this
work, we remark that (\ref{PofQ}) has wider applicability and could
form the basis of further studies of genetic drift in a subdivided
population.  We also note---although we have no use for it here---that
it is possible to write down a similar explicit formula for $Q$ in
terms of $P$ by making use of the identity (\ref{id}).

In the remainder of this work we specialise to the slow migration
limit for the reasons outlined in the introduction.  In particular we
know \citep{Nagylaki80} that if $\mu_{ij}$ is the probability that an
individual sampled at random from deme $i$ had a parent from the
previous generation in deme $j$, the overall population behaves like a
panmictic population with an effective size given by Eq.~(\ref{NeNag})
if one has
\begin{equation}
\lim_{\bar{N} \to \infty} \bar{N} \mu_{ij} = \infty
\end{equation}
where $\bar{N}$ is the mean deme size.  Since we are most interested
in those cases where deviation from panmixia is most likely, and
genetic diversity is maintained for the longest periods of time before
the onset of fixation, we shall insist that
\begin{equation}
\label{mu}
\lim_{\bar{N} \to\infty} \bar{N} \mu_{ij} = m_{ij} < \infty
\end{equation}
holds for all $i$ and $j$.  In this limit it is convenient to work in
rescaled time units, so that that one unit of time corresponds to
$\bar{N}$ generations of the underlying population dynamics.  These
rescaled time units will be in force until the end of this work.
 
In these rescaled time units, the backwards-time hopping of lineages
becomes in the limit $\bar{N}\to\infty$ a continuous-time process in
which a hop from deme $i$ to $j$ occurs with rate $m_{ij}$.
Meanwhile, pairs of lineages collocated in deme $i$ coalesce at a rate
$\bar{N}/\bar{N_i}$ which is assumed to remain finite in the limit
$\bar{N}\to\infty$.  The fact that both of these processes occur on
the same timescale makes calculation of $Q(C;t)$ in general difficult,
since coalescence and migration events are intermingled.  To simplify
the calculations, we shall further assume that all the migration rates
$m_{ij}$ are proportional to some vanishingly small parameter $m$.
This defines an extreme slow migration limit that we shall henceforth
assume is in operation.  Then, we expect that the probability that a
migration event occurs in the time it takes all lineages that are
present in a single deme to coalesce vanishes with $m$, even in the
limit of an infinite system.  This expectation is based on the fact
that the fixation time in an ideal population is of order one (in the
rescaled time units), one can choose $m$ sufficiently small that the
probability that all lineages coalesce before any of them migrate
elsewhere with arbitrarily high probability.  We note that for large,
but finite, populations this assumption has been used previously by
\citet{Takahata91} and formally proved by \citet{Notohara01};
furthermore, \citet{Griffiths84} gives grounds to believe that in an
infinite population, only a finite number of lineages remain at any
nonzero (rescaled) time.  Given that this is the case, the probability
for there to be more than one ancestor in any deme at a time of order
$1/m$ vanishes, and hence all the sums over $c_i$ in (\ref{PAt}) can
be truncated at $c_i=1$.  This yields for the fixation probability as
a function of time the expression
\begin{equation}
\label{Pfix}
P(A; t) = \prod_{i=1}^{L} \sum_{c_i=0}^{1} \chi_i^{c_i} Q(c_1, \ldots,
c_L; t)
\end{equation}
which is the fundamental equation that will be used repeatedly in this
work.  Here, $\chi_i = a_i/N_i$ is the initial fraction of the
individuals in deme $i$ that are mutants and by convention we take
$\chi_i^0 = 1$ whenever $\chi_i=0$.  Any timescales calculated from
this expression will thus be proportional to $1/m$, and it is to be
understood that any times $T$ quoted in the following are believed to
have the property that the combination $mT$ is exact in the limit of
extremely slow migration, $m\to0$.

We remark that the formula for the ultimate probability of fixation,
Eq.~(\ref{Past}), then follows by taking the limit $t\to\infty$ (which
is indicated by the asterisks on $P$ and $Q$).  In this limit $Q(c_1,
\ldots, c_L; t)$ is zero for any state with more than one ancestor;
and converges to $Q_i^\ast$ for the state with a single lineage in
deme $i$.  Since lineages hop from deme $i$ to deme $j$ going backward
in time, this distribution is given by the solution of the balance
equations (\ref{ssmu}) recast in terms of the parameter $m_{ij}$, viz,
\begin{equation}
\label{ss}
\sum_{j \ne i} Q_i^\ast m_{ij} = \sum_{j \ne i} Q_j^\ast m_{ji} \;,
i=1,2, \ldots, L \;.
\end{equation}
It is assumed---as is customary \citep{Nagylaki80,WB97}---that the
stationary solution is unique.  One way that this can be assured is if
it is possible for every deme to be reached from any other through
some sequence of migration events, as is well known from the theory of
Markov chains \citep[see e.g.,][]{GS01}.

\section{Fixation from a random initial condition}
\label{ranic}

Our first application of the formula (\ref{Pfix}) concerns a random
initial condition $A_x$ in which a fraction $x$ of all individuals are
mutants, but are spatially distributed uniformly at random.  The
probability of having $a_1, a_2, \ldots$ mutants in demes $1, 2,
\ldots$ is then given by
\[
\frac{1}{{L\bar{N} \choose xL\bar{N}}} \prod_{i=1}^L { N_i \choose a_i } \;.
\]
The probability of fixation by time $t$ from this random condition is
obtained by summing the right-hand side of (\ref{Pfix}) over all
initial conditions $A$ that have $|A| \equiv \sum_{i} a_i = xL\bar{N}$,
weighted by the previous expression.  In this sum one encounters the
combination
\[
\sum_{A} \prod_{i=1}^{L} \left(\frac{a_i}{N_i}\right)^{c_i} {N_i
\choose a_i} \delta_{|A|, xL\bar{N}} \;.
\]
This can be evaluated by noting that it is the coefficient of
$z^{xL\bar{N}}$ in the power series expansion of
\begin{equation}
\prod_{i=1}^{L} \left[ \frac{1}{N_i} \frac{\partial}{\partial y_i}
\right]^{c_i} (1+zy_i)^{N_i} = \prod_{i=1}^{L} z^{c_i} \left(1 + z
y_i\right)^{N_i-c_i}
\end{equation}
evaluated at $y_1=y_2=\cdots=y_L=1$.  This reveals the fixation
probability $P_r(x;t)$ from a random initial condition to be
\begin{equation}
P(A_x;t) = \prod_{i=1}^{L} \sum_{c_i=0}^{1} \frac{{L\bar{N}-\sum_j c_j
    \choose xL\bar{N}-\sum_j c_j}}{{ L\bar{N} \choose xL\bar{N}}}
    Q(c_1, \ldots, c_L; t) \;.
\end{equation}
Note now that the combinatorial factor appearing in the sum depends
only on the total number $n=\sum_{j}c_j$ of ancestors of the entire
population that remain after going backwards a time $t$ from the
present day, and not their location.  Since, for any finite
$n$, we have
\begin{equation}
\lim_{\bar{N}\to\infty} \frac{{L\bar{N}-n \choose xL\bar{N}-n}}{{
L\bar{N} \choose xL\bar{N}}} = x^n \;,
\end{equation}
we find in the limit of infinite subpopulation sizes (but fixed,
finite deme number $L$) the simple expression
\begin{equation}
P(A_x;t) = \sum_{n=1}^{L} x^n Q(n;t)
\end{equation}
for the fixation probability, in which $Q(n;t)$ is the probability
that the entire population had precisely $n$ ancestors at a time $t$
prior to the present day.  Comparison with Eq.~(\ref{Pfix}) reveals
that this random initial condition gives the same fixation probability
as a homogeneous initial condition, in which every deme contains a
fraction $\chi_i =a_i/N_i = x$ mutants.  We also see that since
$\lim_{t\to\infty} Q(n;t) = \delta_{n,1}$, the ultimate fixation
probability $P^\ast(A_x)=x$: that is, no matter what the migration
rates are, the fixation probability for a randomly distributed set of
mutants is always equal to their initial overall frequency.

Since the probability that fixation occurs in the time interval $[t,
t+\rmd t]$ is $\frac{\rmd}{\rmd t} P(A;t) \rmd t$, the mean time to
fixation, averaged over those realisations where fixation does occur,
is
\begin{equation}
\label{taudef}
\tau(A) = \frac{1}{P^\ast(A)} \int_0^\infty t \frac{\rmd}{\rmd t}
P(A;t) \rmd t \;.
\end{equation}
For the case of the random initial condition,
\begin{equation}
\tau(A_x) = \frac{\sum_{n=1}^{L} x^n \int_0^\infty t \frac{\rmd}{\rmd
t} Q(n;t) \rmd t}{x} \;.
\end{equation}
Noting that in any realisation of the dynamics, the $n$-ancestor state
is entered or exited exactly once (except for the $n=1$ state which is
never exited), this expression can also be written as
\begin{equation}
\label{taurT}
\tau(A_x) = T_1 + \sum_{n=2}^{L} x^{n-1} (T_n - T_{n-1})
\end{equation}
in which $T_n$ is the mean time at which the state with $n$ ancestors
is entered, going backwards in time from a state with one lineage per
deme (i.e., the whole population in the extreme slow migration limit).
We reiterate that this expression, believed not to have appeared
before, is valid for any set of migration rates (at least, in the
extreme slow migration limit) and further remark that when the
coalescence times $T_n$ cannot be obtained analytically, they can be
easily estimated by Monte Carlo sampling of the genealogies.

\section{Fixation in Wright's island model}
\label{wright}

A model for which fixation times can be calculated analytically for
any initial condition is Wright's island model
\citep{Wright31,Maruyama70b,Latter73}.  This provides one case where
variation of fixation time with initial condition can be fully
explored, and comparison made with the result for a random initial
condition via the result of the previous Section.  These new results
provide more detailed information about the island model in the slow
migration limit than the formul\ae\ for the mean and variance of
coalescence times provided by \cite{Takahata91}.

In this model of population subdivision, migration occurs at a uniform
rate between every pair of demes, and all demes have the same size
$N_i=N$.  In order that results for different numbers of demes can be
compared, we insist that mean number of individuals replaced in each
generation is independent of the number of demes $L$.  We therefore
set $m_{ij}=m/(L-1)$ in which $m$ defines an overall timescale as
observed in Section~\ref{prelim}.

Analysis of the model is relatively straightforward because the
statistics of the genealogies are invariant under relabelling of the
demes.  Therefore, the distribution of ancestors $Q(C;t)$ depends
\emph{only} on the number of lineages $n$ that are contained within
$C$ and not their location.  The fixation probability (\ref{Pfix}) can
thus be written as
\begin{equation}
P(A;t) = \sum_{n=1}^{L} \Gamma_n(\chi_1, \ldots, \chi_L) Q(n;t)
\end{equation}
where $Q(n;t)$ is as defined in the previous section and the
coefficients $\Gamma_n(\chi_1, \ldots, \chi_L)$ are proportional to the
elementary symmetric polynomials in $\chi$:
\begin{equation}
\Gamma_n(\chi_1, \ldots, \chi_L) = \frac{1}{{L \choose n}} \sum_{1 \le i_1
< i_2 < \cdots < i_n \le L} \chi_{i_1} \chi_{i_2} \cdots \chi_{i_n}
\;.
\end{equation}
The ultimate fixation probability $P^\ast(A) = \Gamma_1(\chi_1,
\ldots, \chi_L)$, as can be seen by comparing with Eq.~(\ref{Past}).
Using this in the denominator of (\ref{taudef}), we find the mean time
to fixation is
\begin{equation}
\label{tauAT}
\tau(A) = T_1 + \sum_{n=2}^{L} \frac{\Gamma_n(\chi_1, \ldots,
\chi_L)}{\Gamma_1(\chi_1, \ldots, \chi_L)} (T_n - T_{n-1}) \;.
\end{equation}
Note the similarity with Eq.~(\ref{taurT}) given in the previous
Section for the case of a random initial condition.  Here, an
expression in terms of entry times $T_n$ is possible only because all
states with $n$ ancestors are equiprobable at a given time in the
history of the dynamics.

The entry times themselves can be calculated easily because, as in an
ideal population, the rate of coalescence from a state with $n$
lineages to one with $n-1$ lineages is a Poisson process that occurs
with a rate proportional to $n(n-1)$.  In the slow migration limit,
the constant of proportionality is $m/(L-1)$ \citep[see,
e.g.,][]{Takahata91}, whereas in an ideal population with $n\ll N$, it
is $\frac{1}{2}$.  The mean time spent in the $n$ ancestor state is
then simply
\begin{equation}
T_{n-1}-T_{n} = \frac{L-1}{mn(n-1)} \;.
\end{equation}
As the initial condition comprises $n=L$ lineages, we have that
$T_L=0$, and therefore
\begin{equation}
T_1 = \sum_{n=2}^{L} \left(T_{n-1} - T_{n}\right) = \frac{L-1}{m}
\left(1 - \frac{1}{L} \right) \;.
\end{equation}

The mean time to fixation from a random initial condition $A_x$ is
then obtained by using these expressions in (\ref{taurT}).  We find
\begin{eqnarray}
\tau(A_x) &=& \frac{L-1}{m} \left[ \left(1 -
  \frac{1}{L}\right) + \sum_{n=2}^{L} x^{n-1} \left( \frac{1}{n} -
  \frac{1}{n-1} \right) \right] \\
&=& \frac{L-1}{m x} \left[ (1-x) \sum_{n=1}^{L-1} \frac{x^n}{n}
  + \frac{x^L-x}{L} \right] \;.
\end{eqnarray}
When the number of islands $L$ is large, we can expand the term in
square brackets as a series in $1/L$ to obtain
\begin{equation}
\tau(A_x) \sim \frac{L-1}{m} \left[ - \frac{(1-x)\ln(1-x)}{x}
  - \frac{1}{L} + O(L^{-2}) \right] \;,
\end{equation}
which can be compared to (\ref{tideal}), taking into account that in
this expression time is measured in units of $\bar{N}$ generations,
whereas (\ref{tideal}) measures time in generations alone.

Since, as previously described, the random initial condition is
equivalent to a homogeneous initial condition where a fraction $x$
of the individuals in each deme are mutants, the most interesting
comparison is with a highly inhomogeneous initial condition
$\tilde{A}_x$ that has the same number of mutants.  This is attained
by having a fraction $x$ of the demes containing only mutants, and the
rest containing only the wild type.  For such a distribution,
$\Gamma_n(\chi_1, \ldots, \chi_L) = {Lx \choose n} / {L \choose n}$.
We then find that the mean time to fixation is
\begin{equation}
\tau(\tilde{A}_x) = \frac{L-1}{m} \left[ \left( 1 - \frac{1}{L}
      \right) + \frac{1}{x} \sum_{n=2}^{Lx} \frac{{Lx \choose n}}{{L
      \choose n}} \left( \frac{1}{n} - \frac{1}{n-1} \right) \right] \;.
\end{equation}
To simplify this sum, it is convenient to replace the quantity in the
final set of round brackets with $(\frac{1}{n} - \frac{1}{L}) -
(\frac{1}{n-1}-\frac{1}{L})$.  Collecting terms, one eventually finds
after some rearrangement that
\begin{eqnarray}
\tau(\tilde{A}_x) &=& \frac{L-1}{mx} \sum_{n=1}^{Lx} \left[
  \frac{{Lx \choose n}}{{L \choose n}} - \frac{{Lx \choose n+1}}{{L
  \choose n+1}} \right] \left( \frac{1}{n} - \frac{1}{L} \right) \\
&=& \frac{(L-1)(1-x)}{mx} \sum_{n=1}^{Lx} \frac{{Lx \choose n}}{{L
\choose n}} \frac{1}{n}
\end{eqnarray}
in which the second line was obtained from the first by expanding out
the binomial coefficients and further rearrangement.
For large $L$, this expression can be written as
\begin{equation}
\tau(\tilde{A}_x) \sim \frac{L-1}{m} \left[ - \frac{(1-x)\ln(1-x)}{x}
- \frac{1}{2L} + O (L^{-2}) \right] \;.
\end{equation}

We see that both initial conditions yield in the limit $L\to\infty$
the same functional form as the classic result (\ref{tideal}) for an
ideal population, up to a change of timescale.  Taking into account
that (\ref{tideal}) gives the fixation time in terms of a number of
generations, we find the effective population size of a single
panmictic population to be $N_e = \bar{N} (L-1)/2m$; in what follows a
useful measure will be the effective population size $\alpha_e =
N_e/\bar{N}$ relative to the mean size of a single deme, since this
remains finite in the limit $\bar{N}\to\infty$.  This correspondence
between fixation times can be understood by the fact that, as
described above, the coalescence process at the level of demes in the
slow migration limit is precisely the same as that for an ideal
population, albeit on a longer timescale.  It is perhaps interesting
to observe that whilst (\ref{tideal}) was obtained using forward-time
diffusion equations, these new results have been determined entirely
within the backward-time coalescent formalism thus providing an
explicit demonstration of the equivalence of these two complementary
approaches.

For large, but finite $L$, the difference between the fixation time
for the inhomogeneous and homogeneous initial conditions
($\tilde{A}_x$ and $A_x$ respectively) is
\begin{equation}
\tau(\tilde{A}_x) - \tau(A_x) \sim \frac{1}{2m} + O(L^{-1}) \;.
\end{equation}
As this difference is small on the timescale of fixation, which grows
linearly with the number of demes $L$, we suggest that an
inhomogeneous initial condition relaxes quickly to a homogeneous
state.  This state, which has a frequency $x$ of mutants in each deme,
would then persist for a time proportional to $L$.  This accords with
what one observes in a Monte Carlo simulation of the forward-time
population dynamics \citep{BBCM06}.

We conclude this Section by arguing that the homogeneous (or random)
initial condition gives the shortest fixation time of all possible
distributions that have an overall fraction $x$ of mutants in the
population.  To show this, one needs first to impose the constraint
$\sum_i \chi_i = Lx$ by setting $\chi_L = xL - \sum_{i=1}^{L-1}
\chi_i$.  Then, one finds the extremum of $\tau(A)$ by setting all the
derivatives of the right-hand side of (\ref{tauAT}) with respect to
the independent parameters $\chi_1, \chi_2, \ldots, \chi_{L-1}$ to
zero.  It is a straightforward, but tedious, exercise to show that
$\frac{\partial}{\partial \chi_i} \Gamma_n(\chi_1, \ldots, \chi_L)$ is
proportional to $\chi_L -\chi_i$ with a non-negative constant of
proportionality, except for the case $n=1$ where the derivative
vanishes due to the constraint.  All derivatives of (\ref{tauAT}) thus
vanish when all the fractions $\chi_i$ are equal, which corresponds to
the homogeneous initial condition.  Although this demonstrates only
that this is an extremal fixation time, the fact that the fixation
occurs more slowly from an inhomogeneous initial condition (at least
for large $L$) is suggestive that the extremum is a minimum,
particularly since the analysis just outlined also shows that the
point $\chi_i=x$ (for all $i$) is the only extremum in the interior of
the $(L-1)$-dimensional space of independent parameters $\chi_i$.

\section{Fixation from the first mutation event}
\label{singleseed}

For models of migration that have more structure than Wright's island
model, calculation of the fixation time from an arbitrary initial
condition is much more difficult.  We thus specialise to the
comparatively simple case of an initial condition that has a single
mutant in the entire population.  Our approach is to assume that
certain statistical properties of the most recent common ancestor
(MRCA) of the whole population are known, e.g., from an explicit
calculation or Monte Carlo sampling.  Then, starting from
(\ref{taudef}) we derive a new formula for the fixation time from a
single mutation as a function of its location.  We first present this
derivation, and then illustrate its implications through two explicit
and contrasting models of population subdivision.

\subsection{Relation between MRCA statistics and fixation time}
\label{mrca}

If we have an initial condition $A_i$ in which a fraction $\chi$ of
the individuals in deme $i$ are mutants, and all others in the population
are the wild type, we have, for arbitrary $\chi$, the mean fixation
time
\begin{equation}
\label{taui}
\tau_i \equiv \tau(A_i) = \frac{\int_0^\infty t \frac{\rmd}{\rmd t}
  Q_i(t) \rmd t}{Q^\ast_i} \;,
\end{equation}
in which $Q_i(t)$ is the probability that, a time $t$ prior to the
present day, the entire population has a single ancestor that is
located in deme $i$.  To make a connection with the MRCA, we introduce
three quantities: first, the probability density $r_j(t)$ for the
single ancestor state to be entered in deme $j$ at time $t$; second,
the integral of this quantity $R_j^\ast = \int_0^{\infty} r_j(t) \rmd
t$ which gives the total probability that the MRCA is in deme $j$; and
finally $Q_{ji}(\Delta t)$ for the single ancestor of the whole
population to be in deme $i$ a time $\Delta t$ after it was in deme
$j$ going backwards in time.  With these definitions, we then have
that
\begin{equation}
Q_i(t) = \sum_{j=1}^{L} \int_0^t \rmd t' r_j(t') Q_{ji}(t-t') \;.
\end{equation}

The numerator of (\ref{taui}) is then
\begin{multline}
\label{nasty}
\int_0^{\infty} t \frac{\rmd}{\rmd t} Q_i(t) \rmd t=
   \sum_{j=1}^{L} \int_{0}^\infty t \bigg[ r_j(t)
   Q_{ji}(0) + {} \\
\int_0^t \rmd t' r_j(t') \frac{\rmd}{\rmd t}
   Q_{ji}(t-t') \bigg] \rmd t \;.
\end{multline}
The double integral in this expression can be written as
\begin{multline}
\int_0^{\infty} \rmd t' \int_0^{\infty} \rmd t (t'+t) r_j(t')
  \frac{\rmd}{\rmd t} Q_{ji}(j;t) =\\ \int_0^{\infty} \rmd t' t'
  r_j(t') \big[ Q_i^\ast - Q_{ji}(0) \big] + R_j^\ast \int_0^{\infty}
  t \frac{\rmd}{\rmd t} Q_{ji}(t) \rmd t \;.
\end{multline}
Two simplifications now occur: first, the term containing $Q_{ji}(0)$
cancels that in (\ref{nasty}); second, we have that
\begin{equation}
\sum_{j=1}^{L} \int_0^\infty \!t \, r_j(t)\, \rmd t = T_1 \;,
\end{equation}
since the total probability density for the MRCA to be found at time
$t$ is the sum $\sum_j r_j(t)$.  The expression (\ref{nasty})
consequently reduces to
\begin{equation}
\label{nice}
\int_0^{\infty} t \frac{\rmd}{\rmd t} Q_i(t) \rmd t =
Q_i^\ast T_1 +
\sum_{j=1}^{L} R_j^\ast \int_0^{\infty} t \frac{\rmd}{\rmd t}
  Q_{ji}(t) \rmd t \;.
\end{equation}

To attack the remaining integral we diagonalise the $L\times L$ matrix
$\matr{M}$ that has elements
\begin{equation}
[\matr{M}]_{ij} = \left\{ \begin{array}{ll} m_{ij} & i \ne j \\
-\sum_{i \ne j} m_{ij} & i = j \end{array} \right.
\end{equation}
and is the generator of the Markov process the describes the
backward-time hopping in the single-ancestor state.  This matrix has
$L$ eigenvalues, one of which $\lambda_1$ is zero and corresponds to
the stationary state: the left and right eigenvectors have elements
$[\lev{u}{1}]_i = Q^\ast_i$ and $[\rev{v}{1}]_i = 1$ respectively.
The remaining eigenvalues $\lambda_2, \ldots, \lambda_L$ all have a
strictly negative real part, since the stationary state is by
assumption unique.  The left and right eigenvectors satisfy the
biorthogonality relation $\lev{u}{n} \cdot \rev{v}{m} = \delta_{n,m}$.

The probability density $Q_{ji}(t)$ can then be written as
\begin{equation}
Q_{ji}(t) = [\rme^{\matr{M}t}]_{ji} = Q_i^\ast + \sum_{n=2}^{L}
\rme^{\lambda_n t} [\lev{u}{n}]_i [\rev{v}{n}]_j
\end{equation}
in which the stationary solution has been separated out for clarity.
We can now evaluate the integral in (\ref{nice}),
\begin{equation}
\int_0^\infty t \frac{\rmd}{\rmd t} Q_{ji}(t) \rmd t = \sum_{n=2}^{L}
\frac{[\lev{u}{n}]_i [\rev{v}{n}]_j}{\lambda_n} \;,
\end{equation}
which allows us finally to write down an expression for the fixation time
$\tau_i$ originally given by (\ref{taui}).  It reads
\begin{equation}
\label{result}
\tau_i = T_1 + \frac{1}{Q_i^\ast} \sum_{n=2}^{L}
\frac{[\lev{u}{n}]_i}{\lambda_n} \left( \scp{\vec{R}^\ast}{\rev{v}{n}} \right)
\end{equation}
where $\vec{R}^\ast$ is the (row) vector of probabilities for the
location of the MRCA. Thus the problem of calculating the mean time to
fixation from the first mutation event in deme $i$ is reduced to that
of determining the mean time to the MRCA, $T_1$, its spatial
distribution $R^\ast_i$ and the eigenvalues and eigenvectors of the
$L\times L$ matrix of migration rates $\matr{M}$.

Before applying this formula to concrete models, we remark that should
the MRCA distribution $R_i^\ast$ and stationary distribution
$Q_i^\ast$ coincide, $\tau_i=T_1$ for all $i$.  This is because then
$\vec{R}^\ast$ is the zero left eigenvector of $\matr{M}$,
$\rev{u}{0}$, and the scalar product $\scp{\vec{R}^\ast}{\rev{v}{n}}$
vanishes by the biorthogonality of the eigenvectors.  One situation
where this occurs is when any pair of demes can be exchanged without
affecting the dynamics, as in Wright's island model: then both
$R_i^\ast$ and $Q_i^\ast$ are uniform.

In general, the difference between $\tau_i$ and $T_1$ will be
non-zero.  Furthermore, by summing (\ref{result}) weighted by
$Q_i^\ast$, or taking the limit $x\to0$ in Eq.~(\ref{taurT}), the mean
time to fixation from a randomly located mutation is equal to $T_1$.
Hence, $\tau_i$ can be larger or smaller than $T_1$, a result which at
first sight seems counterintuitive, but can be understood from the
fact that one averages over all realisations of the dynamics to find
the mean time to the MRCA, but only over a restricted subset in the
case of fixation.

\subsection{Two example applications}
\label{egs}

We now determine fixation times in two concrete models of population
subdivision undergoing extremely slow migration.  The first is similar
to Wright's island model, in that migration is permitted between every
pair of demes.  However, migration occurs at one of two rates,
depending on whether a pair of demes are considered to belong to the
same, or different, clusters of demes.  The second model has a further
restriction, in that migration between clusters can only occur if one
of those clusters is a special central cluster.

We will compare the data obtained with predictions for the fixation
time from (\ref{tideal}) in the limit $x\to 0$ and using the
asymptotic effective size given by Eq.~(\ref{Ne}).  As has been
established \citep{Slatkin91} the limiting ratios of identity
probabilities appearing in (\ref{Ne}) can be replaced by ratios of
mean coalescence times $Y_{ij}$ for two individuals, one sampled from
cluster $i$ and one from $j$.  Specifically, one obtains
\begin{equation}
\label{NeY}
N_e =  \frac{\sum_{i=1}^{L} \sum_{j=1}^{L}
Q_i^\ast Q_j^\ast Y_{ij}}{\sum_{i=1}^{L}
\frac{1}{N_i} (Q_i^\ast)^2 Y_{ii}} \;.
\end{equation}
In the foregoing we have assumed that, on the timescale of migration,
the rate of coalescence between pairs of lineages located in a single
deme is infinitely fast.  However, we clearly cannot simply set
$Y_{ii}=0$ in the previous expression to obtain an estimate of the
fixation time; instead, we must take into account that coalescence
occurs at a fast, but finite rate.

To this end, let use return to the original time units where there is
a finite mean subpopulation size $\bar{N}$, and the population evolves
in discrete time.  We will take the size of subpopulation $i$ to be
$N_i = \alpha_i \bar{N}$, and migration probabilities $\mu_{ij} = (m
\nu_{ij})/\bar{N}$.  We recall that the extreme slow migration limit
(and a continuous-time dynamics) is reprised by first taking $\bar{N}
\to \infty$ and subsequently $m\to 0$.  The mean coalescence times are
then given by the solution of the set of linear equations
\begin{equation}
\label{Yeqs}
Y_{ij} = 1 + \sum_{k,\ell} \mu_{ik} \mu_{j\ell} \left( 1 -
\frac{\delta_{k, \ell}}{N_k} \right) Y_{k\ell} \;.
\end{equation}

We anticipate that the leading contribution to the coalescence times
$Y_{ii}$ between pairs of lineages in the same deme grows as $\bar{N}$
in the limit $\bar{N}\to\infty$ and $m \to 0$, whilst $Y_{ij}$ grows
as $\bar{N}/m$ in this limit.  We can thus write the \emph{exact}
solutions to (\ref{Yeqs}) as power series in the (small) parameters
$1/\bar{N}$ and $m$:
\begin{eqnarray}
Y_{ii} &=& \bar{N} \left( Y_{ii}^{(0)} + O(m) + O(1/\bar{N})  \right) \\
Y_{ij} &=& \frac{\bar{N}}{m} \left( Y_{ij}^{(0)} + O(m) + O(1/\bar{N}) \right) \quad \mbox{for}
\; i \ne j \;.
\end{eqnarray}
Substituting these expansions into (\ref{NeY}), and taking the limit
$\bar{N}\to\infty$ followed by $m\to0$ yields
\begin{equation}
\label{ae1}
\alpha_e = \lim_{m \to 0} \lim_{\bar{N} \to \infty} \frac{m
  N_e}{\bar{N}} = \frac{\sum_{i \ne j} Q_i^\ast Q_j^\ast
  Y_{ij}^{(0)}}{\sum_i (Q_i^\ast)^2 \frac{1}{\alpha_i}
  Y_{ii}^{(0)}}
\end{equation}
for the effective size of the entire population $\alpha_e$ relative to
the mean subpopulation size $\bar{N}$.  Thus we need only determine
the leading coeffecients in the series expansions.  From (\ref{Yeqs})
we find linear equations satisfied by these coefficients by again
taking the limit $\bar{N}\to\infty$ followed by $m\to0$.  For the case
$i \ne j$, we have
\begin{equation}
\label{leadingY}
Y_{ij}^{(0)} = \frac{1 + \sum_{k\ne i,j} \nu_{ik} Y_{kj}^{(0)} +
  \sum_{\ell \ne i,j} \nu_{j\ell} Y_{i\ell}^{(0)}}{\sum_{k \ne i}
  \nu_{ik} + \sum_{\ell \ne j} \nu_{j\ell}}
\end{equation}
which are precisely the expressions obtained if one approximates the
coalescence process as one that occurs at an infinite rate.  For the
case $i=j$, however, one finds the finite result
\begin{equation}
Y_{ii}^{(0)} = \alpha_i \left( 1 + 2 \sum_{j \ne i} \nu_{ij}
Y_{ij}^{(0)} \right) \;.
\end{equation}
Substituting this expression into (\ref{ae1}) we find the formula
\begin{equation}
\label{alphae}
\alpha_e = \frac{\sum_{i \ne j} Q_i^\ast Q_j^\ast
  Y_{ij}^{(0)}}{\sum_i (Q_i^\ast)^2 \left( 1 + 2 \sum_{j \ne i}
  \nu_{ij} Y_{ij}^{(0)} \right) }
\end{equation}
for the relative effective population size.  Note that this
expression, valid in the slow migration limit, depends only on
fixation times between pairs of lineages in different demes, and the
deme sizes $\alpha_i$ do not enter.  Finally, using (\ref{tideal}) in
the limit $x\to0$, and returning to rescaled time units (i.e., those
that have one unit of time corresponding to $\bar{N}$ generations of
the population dynamics), we arrive at an estimate of the fixation
time $\tau_e$ that behaves as
\begin{equation}
\label{taue}
\tau_e \sim \frac{2 \alpha_e}{m}
\end{equation}
in the limit $m\to 0$, with $\alpha_e$ given by (\ref{alphae}) in
terms of the solutions to (\ref{leadingY}).  It is with this estimate
that we shall compare data for the two different concrete models.

\subsubsection{Two-level model}

Our first concrete model has $\ell$ equal-sized clusters, $n_i=n$.
Every deme receives a fraction $y$ of its migrants from demes within
its own cluster, and the remaining fraction $1-y$ from demes in other
clusters.  So that this fraction is well defined, we will insist that
$n \ge 2$ in all cases.  We will also have the total rate of migration
into deme $i$, $m_i = \sum_j m_{ij}$ equal to the small parameter $m$
in each deme.  These considerations specify the parameters $\nu_{ij}$
appearing in the migration rates $m_{ij} = m \nu_{ij}$ as
\begin{equation}
\nu_{ij} = \left\{ \begin{array}{cc} \nu_s = \frac{y}{n-1} & i,j
  \;\mbox{from same cluster} \\
\nu_d = \frac{1-y}{n(\ell-1)} & i,j \; \mbox{from different clusters}
\end{array} \right. \;.
\end{equation}
We remark that this hierarchical version of the island model has
previously been considered by \cite{SV91}.  It has the special
property that the dynamics are unaffected by exchanging any pair of
demes, and so the distribution $R_i^\ast$ of the MRCA is uniform;
since $\sum_{j\ne i}m_{ij} = \sum_{j\ne i}m_{ji}$ we also have from
Eq.~(\ref{ss}) a uniform stationary distribution $Q_i^\ast$.
Therefore $\tau_i$ does not depend on $i$, and is thus equal to $T_1$.
Note also that the special value $y = (n-1)/(L-1)$ corresponds exactly
to Wright's island model, and for larger values of $y$ one has faster
migration between demes within the same cluster than between those in
different clusters.

\begin{figure}
\begin{center}
\includegraphics[width=0.9\linewidth]{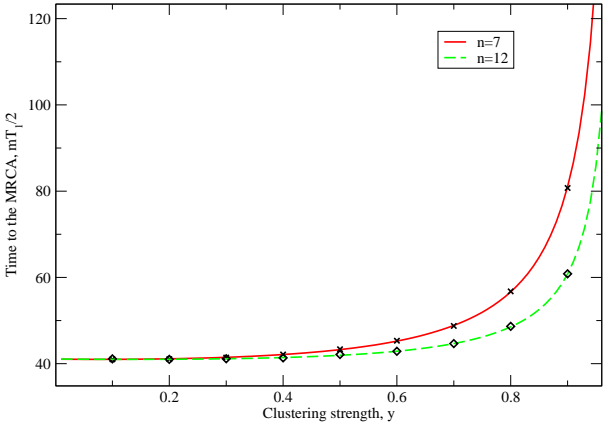}
\end{center}
\caption{\label{compar} Comparison of numerical calculations of the
  time to the MRCA $T_1$ in the two-level model with $L=84$ demes and
  two different cluster sizes, $n=7$ and $12$.  The curved lines show
  data from a numerically exact solution, and the points the
  corresponding averages over $10^5$ genealogies generated by Monte
  Carlo sampling.  The sampling error on the latter are smaller than
  the symbol size.}
\end{figure}

We plot in Fig.~\ref{compar} the combination $mT_1/2$ as a function of
$y$ obtained using two different numerical methods: this combination
then gives an empirical definition of $\alpha_e$ through
Eq.~(\ref{taue}).  The approach is to solve the linear equations
(\ref{Yeqs}) extended to states comprising more than two lineages
\citep[the details of this method are given by, e.g., ][]{Notohara01};
the second method is Monte Carlo sampling of the ancestries.  In turns
out that the former approach is computationally tractable only up to
$L \approx 80$, and so the latter is preferred for larger system
sizes.  At the smaller values of $L$ where both approaches are
possible, we see that the data are---up to numerical
errors---indistinguishable.  Although not particularly evident from
the figure, it turns out that $T_1$, and hence the fixation time, is
always shortest for that value of $y$ that corresponds to Wright's
island model.  One also sees a divergence in $T_1$ as $y\to1$, since
then inter-cluster migration is prohibited.

We now investigate the growth of fixation time with $L$ in the regime
where intra-cluster migration is (at least for sufficiently large $L$)
faster than inter-cluster migration, and compare the effective number
of demes so obtained with the predictions of (\ref{alphae}).  Of the
many possible ways in which the limit $L\to\infty$ can be taken, two
are of particular interest to us.  In the first, the number of
clusters $\ell$ is held fixed whilst the number of demes in each
cluster $n$ goes to infinity; the the second, the cluster size $n$ is
held fixed as more and more clusters are added.

\begin{figure}
\begin{center}
\includegraphics[width=0.9\linewidth]{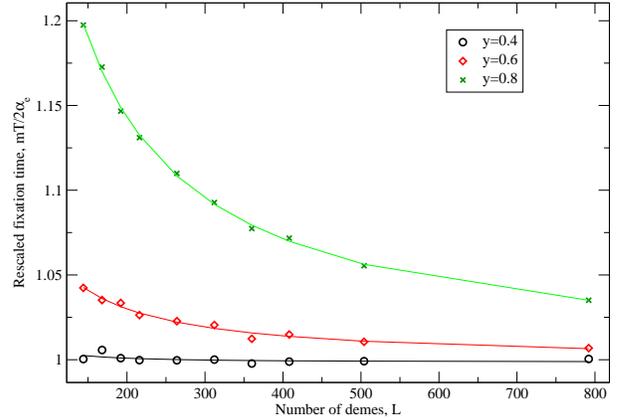}
\end{center}
\caption{\label{twolev_a} Mean time to the MRCA (and hence to fixation
  from a single mutant) normalised by the prediction for the effective
  number of demes for the two-level model as a function of system size
  with fixed cluster number $\ell=12$ and for $y=0.4, 0.6, 0.8$.  The
  points show numerical data, with errors of the same order as symbol
  size.  The solid lines are empirical fits $mT/2\alpha_e = A +
  BL^{-\gamma}$. In all cases $A\approx 1$.}
\end{figure}

\begin{figure}
\begin{center}
\includegraphics[width=0.9\linewidth]{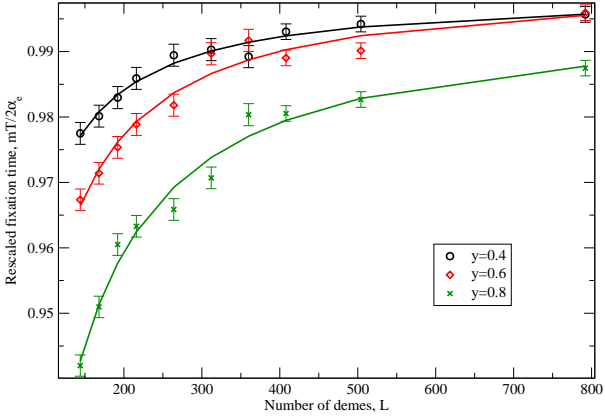}
\end{center}
\caption{\label{twolev_b} As Fig.~\ref{twolev_a} but for the case of
  constant cluster size $n=12$.  Again, the solid lines are fits of
  the form $mT/2\alpha_e = A + BL^{-\gamma}$ with asymptotes $A \approx 1$.}
\end{figure}

Since the steady state is uniform, $Q_i^\ast = \frac{1}{L}$, we find
from Eq.~(\ref{alphae}) that
\begin{equation}
\label{twolevLe}
\alpha_e = \frac{(n-1) Y_s + n(\ell-1) Y_d}{1 + 2y Y_s + 2 (1-y) Y_d}
\end{equation}
where $Y_s=Y_{ij}^{(0)}$ for a pair of demes $i, j$ belonging to the
same cluster, and $Y_d$ the corresponding quantity for two demes from
different clusters.  The equations (\ref{leadingY}) become
\begin{eqnarray}
Y_s &=& \frac{1 + 2n(\ell-1) \nu_d Y_d}{2[ \nu_s + n(\ell-1) \nu_d]}
\\ Y_d &=& \frac{1 + 2(n-1) \nu_d Y_s}{2n \nu_d} \;.
\end{eqnarray}
Substituting the solutions into (\ref{twolevLe}) and taking the limits
of interest one finds that
\begin{equation}
\alpha_e \sim \left\{ \begin{array}{ll}
\frac{L}{2} & n\to\infty,\; \mbox{fixed $\ell$} \\
\frac{L}{2} \left(1 + \frac{y^2}{(n-[1-y])(1-y)}\right) & \ell\to\infty,\; \mbox{fixed $n$}
\end{array} \right.
\;.
\end{equation}
That is, in both cases, the relative effective population size
$\alpha_e$ grows linearly with $L$ asymptotically.  This behaviour is
seen in the numerical data shown in Fig.~\ref{twolev_a} (for constant
$\ell$) and Fig.~\ref{twolev_b} (for constant $n$) in which the
combination $mT_1/2 \alpha_e$ is plotted, with the predictions for
$\alpha_e$ given by the above formul\ae.  Also shown in the figure are
fits to the data $mT_1/2\alpha_e = A + BL^{-\gamma}$, a value of $A
\approx 1$ then indicating a correct prediction for $\alpha_e$.  We
see that in for both cases of fixed cluster number ($\ell=12$) and
fixed cluster size ($n=12$), the asymptotes for three values of the
parameter $y=0.4,0.6$ and $0.8$ are all consistent with a value of
$1$, thus demonstrating the accuracy of the predictions of
(\ref{alphae}) in these instances.

\subsubsection{Hub-and-spoke model}

\begin{figure}
\begin{center}
\includegraphics[width=0.5\linewidth]{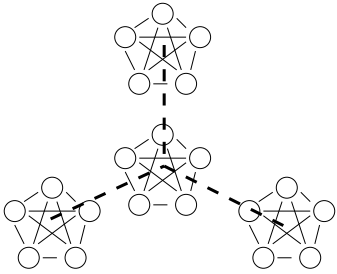}
\end{center}
\caption{\label{hands} Cartoon of the hub-and-spoke migration model.
  The demes are divided into $\ell$ clusters, within which migration
  occurs between every pairs of demes.  Migration between demes lying
  in different clusters is allowed only if one of those demes is in
  the central hub; spoke-to-spoke migration is prohibited.}
\end{figure}

In the second example model, we also divide the demes into $\ell$
clusters of equal size $n\ge 2$, but this time one of the clusters is
a special \emph{hub} deme with migration between clusters permitted
only if one of the two clusters is the hub.  The remaining clusters
thus form \emph{spokes}---see Fig.~\ref{hands}.  This model is
intended to reproduce an effect noticed in social networks, in which
some members of a society interact more widely than others; in a
biological context, one could perhaps interpret this model as
reflecting a continental-archipelago formation, but with a structured
population on the continent.  Either way, we introduce three
independent migration rates: $m_0 = m \nu_0$ for migration between
demes within the same cluster; $m_{hs} = m \nu_{hs} $ for migration
from a spoke deme to a hub deme (going forwards in time); and the rate
$m_{sh} = m \nu_{sh}$ for migration in the opposite direction.

There are at least two ways in which one can make a meaningful
comparison with the results of the previous section.  First, one can
impose a uniform overall rate of immigration $m$, with each deme
receiving a fraction $y$ of immigrants coming from within the cluster,
and the remaining fraction from outside the cluster.  In such a case,
the parameters appearing in the migration rates are
\begin{eqnarray}
\nu_{0} &=& \frac{y}{n-1} \\
\nu_{hs} = \nu_{hs}^{(i)} &=& \frac{1-y}{n(\ell-1)} \\
\nu_{sh} = \nu_{sh}^{(i)}&=& \frac{1-y}{n} \;.
\end{eqnarray}
Alternatively, one can impose a uniform emigration rate, with a
fraction $y$ of migrants from each deme remaining within the cluster,
and the remainder going elsewhere.  This rule is enforced by choosing
$\nu_{hs} = \nu_{hs}^{(e)} = \nu_{sh}^{(i)}$ and $\nu_{sh} =
\nu_{sh}^{(e)} = \nu_{hs}^{(i)}$.  When the symbol $\nu_{hs}$ or
$\nu_{sh}$ appears in expressions below, they are valid for either
rule once the appropriate expressions in terms of $m$, $y$, $n$ and
$\ell$ have been inserted.  Note that the diagonal elements of the
matrix of migration rates will be altered as a consequence of
exchanging the off-diagonal elements so that probability is conserved;
note also that for the two-level model, both rules are equivalent.

One property of the hub-and-spoke model is that the stationary state
satisfies detailed balance, as one can show by using, e.g., a
Kolmogorov criterion \citep{Kelly79}.  This implies that ratios of
stationary probabilities satisfy the relation $Q_i^\ast / Q_j^\ast =
\nu_{ji}/\nu_{ij}$ where $i$ and $j$ label demes.  Introducing
$\tilde{Q}_h^\ast$ for the \emph{total} probability for the single
remaining ancestor of the entire population to reside somewhere in the
hub, and its complement $\tilde{Q}_s^\ast = 1- \tilde{Q}_h^\ast$ for
the ancestor to be somewhere in the spoke, we have that $(\ell-1)
Q_h^\ast/Q_s^\ast = \nu_{sh}/\nu_{hs}$ and so for both rules
\begin{equation}
\tilde{Q}_h^\ast = \frac{\nu_{sh}}{\nu_{sh}+(\ell-1)\nu_{hs}} \;.
\end{equation}
In particular, under uniform immigration one has $\tilde{Q}_h^\ast =
\tilde{Q}_s^\ast = \frac{1}{2}$.  However, since there are (for
$\ell>2$) more spoke demes than hub demes, a mutation occurring in the
hub is $(\ell-1)$ times more likely to fix than one occurring in the
spoke.  Under the uniform emigration rule, the opposite is true, a
mutation somewhere in the spoke being $(\ell-1)$ times more likely to
fix.

A further useful property of the hub-and-spoke model is that the
diagonalisation of the matrix $\matr{M}$ that is needed to calculate
$\tau_i-T_1$ via Eq.~(\ref{result}) can be done analytically. In fact,
only one non-stationary eigenstate contributes.  This state has
\begin{eqnarray}
\lambda_2 &=& - m \left[ n \nu_{sh} + n (\ell-1) \nu_{hs} \right] \\
\lev{u}{2} &=& ( 1, -1) \\
\rev{v}{2} &=& ( \tilde{Q}^\ast_s, -\tilde{Q}^\ast_h)^T
\end{eqnarray}
where the $i^{\rm th}$ element of a column vector notated here as $(h,s)^T$
is $h$ if $i$ corresponds to a hub deme, and $s$ otherwise, whilst the
corresponding element of row vector $(h,s)$ is $h/n$ for a hub
deme, and $s/[n(\ell-1)]$ otherwise.  Defined this way, the scalar
product $(h,s)\cdot(h',s')^T = hh'+ss'$, and the stationary solution
can be expressed as $\lev{u}{1} = (\tilde{Q}^\ast_h,
\tilde{Q}^\ast_s)$ which is orthogonal to $\rev{v}{2}$, as required.
With this notation established, it is easy to show that the row vector
\begin{equation}
\vec{R}^\ast \equiv (\tilde{R}^\ast_h, \tilde{R}^\ast_s) =
\lev{u}{1} + \left( \tilde{R}_h^\ast \tilde{Q}_s^\ast -
\tilde{R}_s^\ast \tilde{Q}_h^\ast \right) \rev{u}{2} \;,
\end{equation}
where $\tilde{R}_h$ and $\tilde{R}_s$ are the total probability that
the MRCA is somewhere in the hub or a spoke respectively.  As a
consequence of this relation, we must have
$\scp{\vec{R}^\ast}{\rev{v}{n}} = 0$ for $n>2$ by the biorthogonality
of the eigenvectors.  Therefore, only one term $n=2$ in the sum in
(\ref{result}) contributes, as claimed.

Evaluating (\ref{result}) for the case of an initial mutation
positioned somewhere in the hub, one finds the mean time for
subsequent fixation to be
\begin{eqnarray}
\tau_h &=& T_1 + \frac{1}{Q_h^\ast}
       \frac{\scp{\vec{R}^\ast}{\lev{v}{2}}}{\lambda_2} \\ &=& T_1 -
       \frac{1}{|\lambda_2|} \left( \frac{R_h^\ast}{Q_h^\ast} - 1
       \right) \;.
\end{eqnarray}
Written this way, it is evident that when the MRCA is more likely to
be in the hub than the single remaining ancestor in the stationary
state, the mean time to fixation from the hub is less than that to the
MRCA.

Under the uniform immigration rule, the mean time to fixation from the
hub is given by
\begin{equation}
\label{tauhi}
\tau_h = T_1 - \frac{2 R_h^\ast - 1}{2m(1-y)} \;.
\end{equation}
Meanwhile, when uniform emigration is enforced,
\begin{equation}
\label{tauhe}
\tau_h = T_1 - \frac{1}{m(1-u)}\left[ (\ell-1)R_h^\ast -
 \frac{1}{(\ell-1)^{-1} + (\ell-1)} \right] \;.
\end{equation}
The corresponding fixation times from a spoke deme can be found using
the fact that $\tilde{Q}_h^\ast \tau_h + \tilde{Q}_s^\ast \tau_s =
T_1$, as was mentioned at the end of Section~\ref{mrca} above.

It remains to find $T_1$ and $R_h^\ast$ numerically; first, however,
we obtain a prediction for the effective number of demes using
Eq.~(\ref{alphae}).  This is a more involved enterprise than for the
two-level model, because there are four distinct ways to sample pairs
of individuals from the population: both from the hub cluster (we
denote this $hh$), both from the same spoke cluster ($ss$), one from
the hub and one from a spoke ($hs$) and two from different spoke
clusters ($s\bar{s}$).  The set of linear equations (\ref{leadingY})
then becomes
\begin{eqnarray}
Y_{hh} &=& \frac{1 + 2 n (\ell-1) \nu_{hs} Y_{hs}}{2 \nu_0 + 2n(\ell-1)
  \nu_{hs}} \\
Y_{ss} &=& \frac{1 + 2 n \nu_{sh} Y_{hs}}{2 \nu_0 + 2 n \nu_{sh}} \\
Y_{hs} &=& \frac{1}{n \nu_{sh} + n (\ell-1) \nu_{hs}} \big[ 1 + (n-1)
  \nu_{hs} Y_{ss} + {} \nonumber\\
&& \qquad n (\ell-2) \nu_{hs} Y_{s\bar{s}} + (n-1) \nu_{sh} Y_{hh} \big] \\
Y_{s\bar{s}} &=& \frac{1 + 2 n \nu_{sh} Y_{hs}}{2 n \nu_{sh}} \;.
\end{eqnarray}
These can be readily solved using (for example) a computer algebra
package.  The formula for relative effective size (\ref{alphae}) takes
the form
\begin{equation}
\textstyle
\alpha_e = \frac{(\tilde{Q}_s^\ast)^2 [(n-1) Y_{ss} + n (\ell-2) Y_{s\bar{s}}] +
(\ell-1) \tilde{Q}_h^\ast [ (n-1)\tilde{Q}_h^\ast Y_{hh} + 2n \tilde{Q}_s^\ast
Y_{hs}]}{(\tilde{Q}_s^\ast)^2 [1 + 2(n-1)\nu_0 Y_{ss} + 2 n \nu_{sh}
    Y_{hs}] + (\ell-1) (\tilde{Q}_h^\ast)^2 [1 + 2(n-1) \nu_0 Y_{hh} +
  2n(\ell-1) \nu_{hs} Y_{hs}]} \;.
\end{equation}
We shall not present the full expression for $\alpha_e$ here, as it is
rather complicated and unrevealing.  Instead, we shall show only how
$\alpha_e$ behaves asymptotically for large $L$, under the two
immigration rules, and for the two contrasting limits of infinite
system size discussed for the two-level model.

\begin{figure}
\begin{center}
\includegraphics[width=0.9\linewidth]{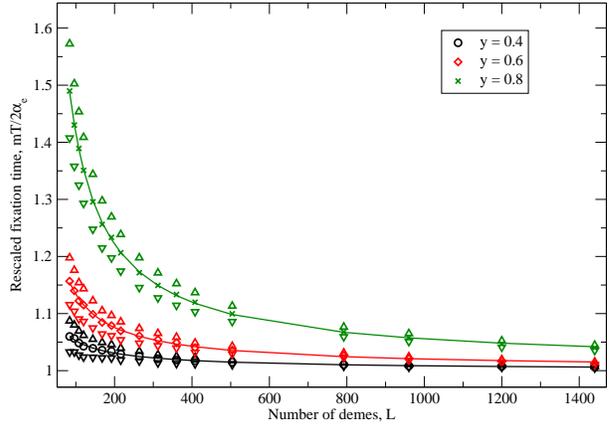}
\end{center}
\caption{\label{hubspkl_a} Mean time to fixation and the MRCA,
  normalised by the relative effective population size $\alpha_e$, in
  the hub-and-spoke model under the uniform immigration rule for
  $y=0.4,0.6,0.8$ and increasing deme number $L$ with fixed cluster
  number $\ell=12$.  The upward- and downward-pointing triangles show
  fixation times from the spoke $\tau_s$ and hub $\tau_h$
  respectively, whilst the points in between are the MRCA times.  The
  solid lines show fits to the latter of the form $mT_1/2\alpha_e = A +
  BL^{-\gamma}$; in all three cases the asymptote $A\approx 1$.}
\end{figure}

\begin{figure}
\begin{center}
\includegraphics[width=0.9\linewidth]{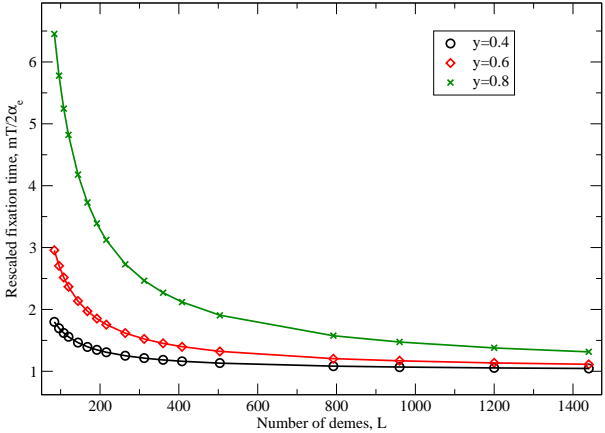}
\end{center}
\caption{\label{hubspkl_b} As Fig.~\ref{hubspkl_a}, but for the
  uniform emigration rule.  In this case, the mean fixation times from
  both hub and spoke are indistinguishable from the mean time to the
  MRCA, so only the latter has been plotted.  The solid lines show
  fits of the form $mT_1/2\alpha_e = A + BL^{-\gamma}$, and show asymptotes
  $A\approx 1$ in all cases.}
\end{figure}

In the case where the number of clusters $\ell$ is held fixed, and the
number of demes within them $n$ increased, the predicted $\alpha_e$
increases linearly with $L$ under both the uniform immigration and
emigration rule.  Specifically one has for uniform immigration
\begin{equation}
\label{LeL1}
\alpha_e \sim \frac{2(\ell-1)}{\ell^2} L
\end{equation}
and for uniform emigration
\begin{equation}
\alpha_e \sim \frac{[ 1 + (\ell-1)^2]^2}{2 \ell^2 [ (\ell-1) +
    (\ell-2)^2 y]} L \;.
\end{equation}
Simulation data for case of fixed $\ell=12$ are shown in
Fig.~\ref{hubspkl_a} for uniform immigration and in
Fig.~\ref{hubspkl_b} for uniform emigration.  In both figures the
combination $mT/2\alpha_e$ is plotted for three different values of
$y$ ($y=0.4,0.6,0.8$). In all cases fits to the data of the form $A+B
L^{-\gamma}$ reveal an asymptote $A$ consistent with unity; that is,
the data show an asymptotic linear growth with the slope predicted by
the effective size formula (\ref{alphae}).  In both cases, the
relative variation of fixation time with the location of the first
mutation vanishes in the limit $L\to\infty$.  That this should be the
case can be seen from the formul\ae\ in Eq.~(\ref{tauhi}) and
(\ref{tauhe}); when $\ell$ is fixed, the correction term is bounded
whilst $T_1$ grows linearly $L$.

\begin{figure}
\begin{center}
\includegraphics[width=0.9\linewidth]{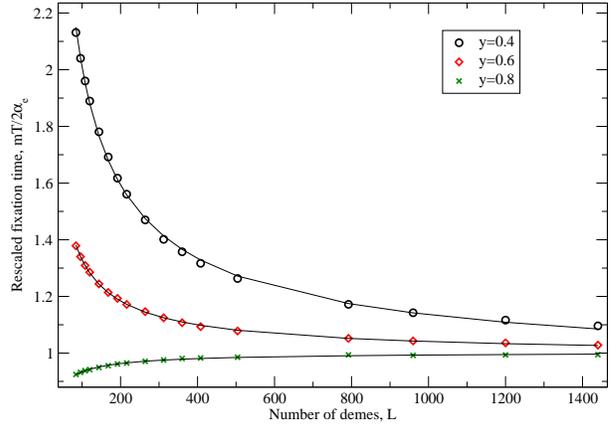}
\end{center}
\caption{\label{hubspkn_a} As Fig.~\ref{hubspkl_a} but for the uniform
  emigration rule and increasing $L$ as the cluster size is held fixed
  at $n=12$.  Here the points show numerical estimates of the mean
  time to the MRCA for $y=0.4, 0.6, 0.8$; again the mean fixation
  times from hub and spoke are indistinguishable from these data. The
  solid lines are fits of the form $mT/2\alpha_e = A + BL^{-\gamma}$, with
  $A\approx 1$ in all three cases.}
\end{figure}

\begin{figure}
\begin{center}
\includegraphics[width=0.9\linewidth]{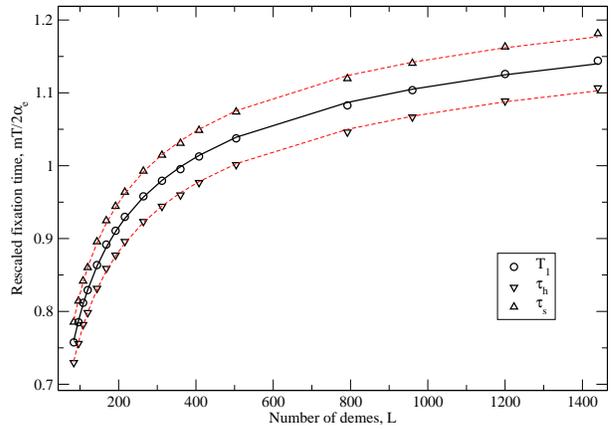}
\end{center}
\caption{\label{hubspkn_b} As Fig.~\ref{hubspkl_a}, but for the
  uniform immigration rule and increasing $L$ as the cluster size is
  held fixed at $n=12$.  Here, the points show numerical data for the
  case $y=0.8$; those for other $y$ values are similar and have been
  omitted for clarity.  The solid lines are fits of the form $mT/2\alpha_e =
  A + BL^{-\gamma}$.}
\end{figure}

A scaling of the fixation time that is nonlinear in $L$ is seen when
the number of demes is increased by adding more and more spokes of
fixed size $n$.  In this limit, the prediction for $\alpha_e$ asymptotes to
a \emph{constant}
\begin{equation}
\label{const}
\alpha_e \sim 2 \frac{(n-1)^2}{n-(1-y)} + \frac{5}{4} \frac{1}{1-y}
\end{equation}
under the uniform immigration rule, but increases \emph{quadratically}
with $L$ as
\begin{equation}
\alpha_e \sim \frac{1}{2n^2(1-y)} L^2
\end{equation}
as $L\to\infty$.  Plotting the quantity $mT_1/2\alpha_e$ for the case
$n=12$ suggests that this asymptotic quadratic growth is correctly
predicted, shown by the approach to a constant value as $L\to\infty$
in Fig.~\ref{hubspkn_a}.  Again fits to the data suggest asymptotes
consistent with unity, showing the accuracy of the prediction given by
the effective relative population size (\ref{alphae}).  In
Fig.~\ref{hubspkn_b}, data for the uniform immigration rule with
constant cluster size $n=12$ are shown for the case $y=0.8$ (data for
other $y$ values show similar behaviour, but have been omitted for
clarity).  Even at the system sizes shown, the fixation time is still
growing with $L$.  However, the data are suggestive that eventually
the fixation time will saturate to a constant as predicted by
Eq.~(\ref{alphae}).  First, a fit to the data of the form
$A+BL^{-\gamma}$ suggests an asymptote of $A\approx 1.38$, that is,
that the fixation time is bounded by a constant but one that is
approximately $40\%$ larger than that predicted by (\ref{alphae}).
(The asymptote for smaller values of $y$ is also underestimated by
(\ref{alphae}), but not to such a great extent: the numerical data
exceed the prediction by about $10\%$ for $y=0.4$ and about $20\%$ for
$y=0.6$).  Further evidence for the fixation time remaining bounded in
the limit $L\to\infty$ is provided by the fact that---uniquely among
the models and limits considered---the relative differences between
the mean times to fixation and the MRCA remain finite in the limit
$L\to\infty$.  As noted above, the absolute differences are bounded
and so relative differences may remain finite only if the fixation
time saturates.

\section{Discussion and conclusion}
\label{discon}

The aim of this work was to develop a better understanding of the
effects of population subdivision on fixation under neutral genetic
drift.  This was achieved by exploiting a connection between forward-
and backward-time properties of neutral genetic drift which admitted
the derivation of a number of new results for fixation properties in
subdivided populations that are exact in the limit of extemely slow
migration.  Since the underlying motivation was partly to assess the
viability of genetic drift as a mechanism for propagating a social
change, we were particularly interested in establishing how fixation
times grow with the number of demes.  Furthermore, the existence of
historical data requires knowledge of variation of fixation times with
the initial distribution of mutants.

We address the matter of variation first.  It is clear from the
elementary considerations leading to Eq.~(\ref{Past}) that fixation
probability can vary greatly with the initial location of mutants.
For example, if one has in the hub-and-spoke model a mutation which is
known to be located in the hub, it is $(\ell-1)$ times more likely to
invade the whole population than the same mutation occurring in a
spoke, as long as the uniform immigration rule is applied.  Thus, if
the number of spokes is very large, one can approach near certainty
for a mutation to fix by genetic drift alone.  However, this is
balanced by the fact that there are $(\ell-1)$ fewer hub demes than
spoke demes, so a mutation occurring at a random location is as
unlikely to fix in this subdivided population as any other with the
same overall size.  We contrast again with the findings of
\citet{LHN05} which showed almost certain fixation from a random
initial condition driven by selection in a spatially structured
population.  Such behaviour can also arise without the need for
selection as long as one is able to position the initial mutations
strategically, as would be the case when one is constrained by
historical data.

Conversely, we observed only minor variation with the initial
condition in the mean fixation time, averaged over those realisations
of the dynamics where fixation actually occurs.  This was seen both in
new exact results for Wright's island model (Section~\ref{wright}) and in
numerical data for clustered models (Section~\ref{egs}).  In fact, in
all cases the relative magnitude of variation vanishes in the limit of
an infinite system, except possibly in the case where the fixation
time remained constant in that limit.  However, even there, the
variation was of the order of a few percent.  This lack of variation
has two practical benefits.  First, one may as well approximate any
historical data by a random initial condition with the same overall
frequency of mutants, and use Eq.~(\ref{taurT}) to calculate fixation
times from the coalescence times $T_n$ which are easy to obtain
numerically.  We remark that this formula can easily be generalised to
find higher cumulants of the fixation time distribution.  The second
benefit is that if one notices considerable variation of fixation
times in historical data, it may then be possible to rule out genetic
drift as a propagation mechanism as a consequence.  Finally, on the
subject of variation, a curiosity that emerged from the exact analysis
of Wright's island model was evidence for mean time to fixation to be
minimised (subject to a constraint of a fixed overall mutant
frequency) by a random initial condition.  It would be interesting to
show this more rigorously, and to see if this is also the case for a
wider class of models.

By considering the fate of a single mutation in models that had demes
grouped into tightly-knit clusters, we established three different
growth laws for the fixation time under various conditions: linear (as
in Wright's island model), quadratic and approach to a constant.  The
latter scenario is, of course, the most intriguing, since then one has
a population-level change occurring in an infinite population in a
finite time through genetic drift alone.  The origin of this
phenomenon lies in the nonconservative nature of the migration process
in place: the total number of individuals entering the spokes from the
hub under the constant immigration rule vastly exceeds the number
entering the hub.  This has the consequence that, as one goes backward
in time, the probability of finding pair of lineages in a vanishingly
small region of space (the hub) remains finite as the number of demes
is increased; thus their coalescence, and hence fixation, can occur in
a finite time.  By contrast, under the uniform emigration rule, the
probability that lineages are found in separate spokes is finite, and
one must wait a long time until they are both present in the same
deme: this drives the superlinear increase in fixation time.  Although
in a biological context the hub-and-spoke model may be of limited
application, one can think of this enhancement of offspring number as
a form of spatially-dependent selection and it would perhaps be of
interest to see if a similar effect is evident in potentially more
realistic situations where the deme sizes are small and their number
restricted by topological considerations.

On the other hand, in the cultural context of language change where
migration corresponds to a speaker retaining a record of another's
utterances, there is no particular reason to assume conservative
migration.  In fact, a uniform immigration rule as implemented in this
work arises rather naturally, as it corresponds to every speaker
dividing the same amount of attention equally between each person she
listens to.  Furthermore, the hub-and-spoke model probably better
reflects the nature of social interactions, in which some members of
society have more long-range connections than others.  It is unclear,
however, if the phenomenon of a finite fixation time will be seen for
infinite populations on more realistic social networks.  We are
currently investigating this possibility, and results will be reported
in due course.

Finally, we compared results from simulations with predictions from a
formula for effective population size, Eq.~(\ref{Ne}) whose general
form was given by \citet{Rousset04} and that was specialised here to
the extreme slow migration limit in which all migration rates are
inversely proportional to the deme size and have a vanishingly small
coefficient.  The resulting formula, (\ref{alphae}), was found to give
precise predictions for the fixation time in the limit of an infinite
number of demes for all models considered, except when the fixation
time appears to be bounded by a constant; here, the predicted
asymptote was seen to be exceeded by as much as $40\%$.  This would
suggest that in this particular case, the fixation time is not well
characterised by the asymptotic coalescence rate that appears on the
right-hand side of Eq.~(\ref{Ne}).  To explore this possibility, it is
worth considering the extreme case of two demes with arbitrary
migration rates $m_{12}=m \nu_{12}$, $m_{21} = m \nu_{21}$ in the
extreme slow migration limit $m\to0$.  If one starts with one lineage
per deme in this model, the rate of coalescence between these lineages
is $c = m_{12}+m_{21}$, since that is the rate at which one of the
lineages hops from one deme to another, at which time a coalescence
immediately takes place (as least in the limit $m\to0$).  By contrast,
the mean asymptotic rate of coalesence given by the asmyptotic
effective size formula (\ref{alphae}) is $c/ [2 Q_1^\ast
(1-Q_1^\ast)]$ where $Q_1^\ast = m_{21}/(m_{12}+m_{21})$.  For any
choice of migration rates, the true rate of coalescence between the
last pair of lineages is overestimated by a least a factor of $2$, and
hence the effective population size underestimated as was seen for the
hub-and-spoke model.  This two-demes model provides an extreme example
of a case where all coalesences occur before the onset of the
asymptotic regime.  We believe this is also what is happening in the
hub-and-spoke model, and could further explain why this is the only
case in this work in which any variation of fixation time with initial
condition was observed.  Through more careful considerations of the
relevant coalescence events \citep{Rousset04}, one anticipates that
better predictions for the fixation time can be obtained.

We end by remarking that if one is only interested in the general
scaling of the fixation time with the number of demes (and not in
precise estimates of the coefficients), the simple formula of
\citet{Nagylaki80}, $N_e^{-1}=\sum_i (Q_i^\ast)^2/N_i$, is much easier
to apply than the full expression (\ref{Ne}) and can be used as long
as lineages are well mixed by the dynamics and fluctuation effects are
unimportant.  An example of a model in which this simple formula gives
the wrong prediction for scaling is the stepping-stone model in two
dimensions and less \citep{Slatkin91,Rousset97,CD02}.  It is also
perhaps worth noting that the stepping-stone model in the
slow-migration limit $m\to0$ is equivalent to the particle reaction
system $A+A\to A$ that has been of interest to physicists
\citep{Peliti86,benAvraham98}.  The methods that have been employed in
this context allow, in principle, the distribution of ancestors
$Q(C|D;t)$ appearing in the general relation (\ref{PofQ}) to be
calculated for the stepping stone model in one dimension.  It would be
interesting to exploit this connection to obtain further new results
for the properties of fixation in subdivided populations.

\section*{Acknowledgements}

I thank Jonathan Coe for introducing me to the coalescent and Nick
Barton, Gareth Baxter, Bill Croft and Alan McKane for comments on the
manuscript.  I also thank the Royal Society of Edinburgh for the
support of a Personal Research Fellowship, and the Isaac Newton
Institute of Mathematical Sciences for hospitality during the
completion of part of this work.

\end{document}